# DNA-assembled advanced plasmonic architectures

Na Liu[1,2*] and Tim Liedl[3*]

[1]Max Planck Institute for Intelligent Systems, Heisenbergstrasse 3, D-70569 Stuttgart, Germany
[2]Kirchhoff Institute for Physics, University of Heidelberg, Im Neuenheimer Feld 227, D-69120, Heidelberg, Germany
[3]Fakultät für Physik and Center for Nanoscience, Ludwig-Maximilians-Universität, Geschwister-Scholl-Platz 1, 80539 München, Germany

## ABSTRACT:

The interaction between light and matter can be controlled efficiently by structuring materials at a length scale shorter than the wavelength of interest. With the goal to build optical devices that operate at the nanoscale, plasmonics has established itself as a discipline, where near-field effects of electromagnetic waves created in the vicinity of metallic surfaces can give rise to a variety of novel phenomena and fascinating applications. As research on plamonics has emerged from the optics and solid-state communities, most laboratories employ top-down lithography to implement their nanophotonic designs. In this review, we discuss the recent, successful efforts of employing self-assembled DNA nanostructures as scaffolds for creating advanced plasmonic architectures. DNA self-assembly exploits the base-pairing specificity of nucleic acid sequences and allows for the nanometer-precise organization of organic molecules but also for the arrangement of inorganic particles in space. Bottom-up self-assembly thus bypasses many of the limitations of conventional fabrication methods. As a consequence, powerful tools such as DNA origami have pushed the boundaries of nanophotonics and new ways of thinking about plasmonic designs are on the rise.

CONTENTS

## 1. INTRODUCTION

The key component of plasmonics is metals. When light interacts with a metal nanoparticle, collective oscillations of conduction electrons known as particle plasmons are excited. As two metal nanoparticles are brought into proximity, their plasmon oscillations couple to each other, resulting in altered optical behavior that is highly distance dependent. [1-3] This arises from the fact that the plasmon resonance of a metal nanoparticle is influenced by other nanoparticles in its immediate environment. Such coupling effects have a profound impact on plasmonic assemblies for achieving novel phenomena and properties. [4-7]

A prerequisite to build advanced plasmonic architectures is the ability to precisely control the arrangement of metal nanoparticles in space. To this end, DNA represents an ideal construction material owing to its unique sequence specificity.[8] The almost infinite possibilities of designing individual sequences, which can be programmed to interact with each other, allow for an extremely rich versatility of DNA nanostructures.[9,10] In general, DNA nanotechnology can be divided into two categories: structural and dynamic DNA nanotechnology. The former is often used to construct static two-dimensional (2D) and three-dimensional (3D) nano-objects of varying geometries and sizes,[11] while the latter affords a remarkable capability to achieve dynamic nano-devices with desired reconfigurability[12,13] and to perform computational tasks. [14-16]

Over the last two decades, the utilization of DNA nanotechnology in plasmonics has led to unprecedented and unexpected novel research directions, which are mutually advantageous for both fields. When DNA nanotechnology meets plasmonics, it does not merely enable the realization of fascinating plasmonic architectures that were impossible to build before, [17,18] but also brings about tailored optical functionalities that are not possessed by DNA structures alone. [19,20] Thanks to the everlasting research efforts in this inter-disciplinary field, DNA nanotechnology-enabled plasmonics has significantly flourished and rendered possible a variety of interesting applications in biomolecular sensing, [21-26] surface-enhanced Raman and fluorescence spectroscopy, [27-40] diffraction-limited optics, [41], *etc*. In particular, the DNA route to build plasmonic architectures holds distinct advantages over top-down techniques due to its inherent molecular recognition and thus unprecedented nanoscale precision through well-controlled self-assembly processes. It also allows for large-scale and even bulk production of plasmonic nanostructures in a highly parallel manner. Importantly, it opens a new research pursuit towards reconfigurable and autonomous plasmonic nanosystems, which are beyond the state of the art of top-down nanofabrication techniques.

In this review, we aim to provide an overview of the recent progress in DNA-assembled advanced plasmonic architectures. It is noteworthy that we will focus on discrete, non-periodic plasmonic nanostructures as extensive reviews are available on periodic nanostructures such as plasmonic chains, arrays, and lattices assembled by DNA. [17,42] We will first describe the general experimental strategies that have been established for fabricating DNA-assembled plasmonic nanostructures. Then, we will introduce a diverse set of examples according to their characteristic optical properties and functionalities. Subsequently, we will discuss the inevitable evolution from static to dynamic plasmonic systems along with the fast development of this interdisciplinary field. Finally, possible future directions and perspectives on the remaining challenges and open opportunities will be elucidated.

## 2. FABRICATION OF DNA-ASSEMBLED PLASMONIC NANOSTRUCTURES

A general strategy to fabricate DNA-assembled plasmonic nanostructures relies on functionalization of metal nanoparticles, for example functionalization of gold nanoparticles (AuNPs) with thiol-modified single-stranded DNA (ssDNA) as programmable linkers (see Fig. 1a). [43-48] Unbound DNA can be removed by centrifugation and resuspension. For the formation of dimers, AuNPs modified with complementary ssDNA are added together for incubation. Through Watson-Crick base-pairing, AuNP dimers can be formed and subsequently purified from monomers as well as higher-order aggregates by gel electrophoresis. [44] Other than dimers, a variety of structures including homo/hetero dimers and trimers, [49-51] [28,52] tetramers, [53] [54] multimers, [55,56], chains, [57-59] lattices [60] [61] [62] *etc.* can also be achieved. Nevertheless, due to lack of structural rigidity, positioning of particles within individual assemblies cannot be accurately controlled. Also, due to lack of spatially directed organization, particle aggregates rather than discrete structures are often formed and still present after gel electrophoresis. To avoid such problems, a number of innovative protocols have been carried out to anisotropically functionalize AuNPs with DNA, *e.g.* through the use of Janus particles[63] or geometric restrictions imposed by a solid substrate or other particles. [64,65] Noteworthy, Chad Mirkin and others have developed a variety of crystalline assemblies using DNA-functionalized nanoparticles. [42,66-69] Most of these examples will not be discussed here and we recommend the reader to refer to the existing literature.

An alternative strategy to realize DNA-assembled plasmonic nanostructures is based on the DNA origami technique (see Fig. 1b). The concept of DNA origami was introduced by Rothemund in 2006. [70] The process involves the folding of a long scaffold ssDNA strand – usually derived from the M13 phage genome – by hundreds of short staple strands into arbitrary 2D and 3D shapes. [71-75] Owing to the high rigidity[76,77] and addressability[78] of DNA origami, this approach is ideally suited to precisely organize metal nanoparticles in space for constructing plasmonic architectures with well-defined configurations. Metal nanoparticles functionalized with single or multiple DNA linkers can be assembled at designated binding sites through hybridization with their complementary DNA strands extended from DNA origami. [46,47,79] In particular, DNA origami offers positioning accuracy of a few nm, giving rise to superior molecular templates with nanoscale addressability. After ten years of research efforts, a multitude of plasmonic architectures with different structural complexities and tailored optical functionalities have been accomplished using the DNA origami technique.

## 3. DNA-ASSEMBLED STATIC PLASMONIC NANOSTRUCTURES

DNA-assembled static plasmonic nanostructures refer to DNA-assembled plasmonic nanostructures, which can yield well-defined optical properties, but do not exhibit dynamic behavior nor possess structural reconfigurability. Through specific spatial arrangements, such plasmonic nanostructures can exhibit interesting optical phenomena including plasmon hybridization, [50,80-83] Fano resonances, [84-86] magnetic resonances, [87] chiral response, [19,20,88-90] [91] [92] *etc*. They may also serve as powerful plasmonic platforms for surface-enhanced Raman [29] [30] [31] [32] [93] [33] [34] and fluorescence spectroscopy [36] [35] [37,41] to understand fundamental light-matter interaction on the nanoscale.

### 3.1 Plasmon Hybridization

Plasmon hybridization is a theory developed by Nordlander to interpret coupling effects in complex plasmonic nanostructures. [80,81] It states that plasmons in assemblies of metal nanoparticles bear a direct resemblance to electrons in quantum molecular orbitals. Plasmons excited in adjacent metal nanoparticles interact, mix, and hybridize just like the electronic wave functions of atomic and molecular orbitals. The simplest case is plasmon hybridization in a dimer of AuNPs as shown in Fig. 2a. The near-field coupling between the two AuNPs in close proximity can result in the formation of bonding and antibonding modes, respectively. The antibonding mode is located at a higher frequency arising from its larger net dipole moment, when compared to the bonding mode.

Alivisatos *et al*. first demonstrated the realization of discrete AuNP assemblies through DNA-monofunctionalization (see Fig. 2a). [44] AuNPs modified with ssDNA in a 1:1 ratio were achieved by taking advantage of the restricted surface-areas of the 1.4 nm particles. After monofunctionalization, the AuNPs were assembled with secondary short scaffolding strands into homodimers and homotrimers through Watson-Crick base-pairing followed by a gel electrophoresis purification step. Later, the same group reported the assembly of heterodimers and heterotrimers with a similar strategy using 5 nm and 10 nm AuNPs. [49] The optical characterization of the 10 nm homodimers was carried out using UV/Vis spectroscopy. A decrease in absorbance, spectral broadening, and a spectral shift to longer wavelengths were observed for the homodimers with respect to the monomers. These results exemplified the first experimental proof of plasmon hybridization in DNA-assembled plasmonic nanostructures.

The critical parameters that determine the optical properties of a complex plasmonic nanostructure include the sizes, shapes, material compositions of the individual nanoparticles, the interparticle spacing, as well as the overall configuration in space. Due to technical constraints, AuNPs utilized in the early work on DNA-assembled plasmonic nanostructures were rather small ($\leq$ 10 nm) and the interparticle distances were often as large as or even larger than their actual sizes. The resulting small dipole moments and weak coupling strengths thus led to indiscernible spectral changes, when compared to the spectra of the individual AuNPs. This obstacle was overcome only later, thanks to successive methodological improvements in this field. One of the remarkable advancements was achieved by Sönnichsen *et al.*, who demonstrated the first plasmon rulers composed of a 40 nm AuNP dimer or a silver (Ag) nanoparticle dimer, which exhibited distinct plasmon hybridization effect. [22] This landmark work will be discussed later in the section of DNA-assembled dynamic plasmonic nanostructures as it represents one of the very first dynamic plasmonic systems enabled by DNA nanotechnology.

Other than homodimers, heterodimers are also of great interest due to their complex optical response, offering rich information and profound insights into plasmonic coupling mechanisms. Asymmetries are often introduced by applying two particles of different sizes or shapes. As a matter of fact, asymmetry can also be created using different material compositions. As shown in Fig. 2b, Au-Ag heterodimers were assembled using DNA by Sheikholeslami *et al*. [50] Thiolated ssDNA molecules were conjugated on the Au (40 nm) and Ag (30 nm) particles, followed by passivation using thiolated polyethylene glycol to provide enhanced stability. The Au and Ag particles were then mixed in a 1:1 ratio and allowed to hybridize overnight. Optical measurements with polarized scattering spectroscopy revealed the excitation of both the bonding and antibonding plasmonic modes. Interestingly, the antibonding modes were red shifted relative to the resonance of the Ag particle. This anomalous shift was due to the coupling between the Ag particle resonance and the quasi-continuum of the interband transitions in Au, which did not occur in the cases of single-composition dimers presented in the same paper.

The fabrication of Au-Ag heterodimers could also be realized by utilization of DNA origami as shown in Fig. 2c. [82] The rigidity of DNA origami allowed for less variability in the interparticle spacing than the aforementioned cases. 40 nm Ag and 40 nm Au particles were assembled into a heterodimer on a DNA origami platform with a sub-5 nm gap. The different material compositions were verified using scanning electron microscopy and energy-dispersive X-ray characterization. Importantly, the authors also developed a new interparticle-spacing-dependent coupling model for heterodimers, very useful for controlling symmetry breaking in collective plasmon modes. Their study confirmed that direct plasmonic coupling in the Au-Ag dimer could be observed in both experiment and theory only for small gap sizes, as it required the Ag dipolar mode frequency to drop below the frequency of the Au interband transitions.

By placing a Ag NP between two remote AuNPs, Roller *et al*. introduced a concept for ultrafast and low-dissipative transfer energy. [94] ( please cite Toppari et al., ACS Nano, 2013, 7 (2), pp 1291–1298)With its excitation energy level being too high to be occupied, the Ag particle here served as a virtual transmitting state during the passage. As a result, no heat was generated in the Ag particle, while the two Au particles were still perfectly coupled. This behavior could be described by classical simulations and with a quantum mechanical model, both reproducing the experimental results. The remarkably short plasmon transfer times of the non-dissipative plasmon passage were on the femtosecond scale, opening the door towards a new class of plasmonic wave-guides.

**3.2 Plasmonic Fano Resonances**

Fano interference was originally studied in atomic and quantum mechanical systems. [84,86,95-97] In fact, it is also ubiquitous in classical systems, generally when energy transfer from an initial state to a final state takes place *via* two pathways that destructively interfere. Plasmonic systems can also support Fano interference, which often occurs in assemblies composed of closely-packed metal nanoparticles. Such Fano phenomena promise many useful applications in optical sensing, switching, nonlinear, and slow-light devices.

Plasmonic Fano interference in DNA-assembled particle clusters was first demonstrated by Fan *et al*. (see Fig. 3a). [98] Au nanospheres (74 nm) and slightly larger Au nanoshells (62.5 and 92.5 nm for the inner and outer shells, respectively) were functionalized separately with DNA strands that were complementary. They were mixed and incubated together to first form loosely bound structures. The clusters were then compressed into closely-packed pentamers through a drying process on a hydrophilic substrate. Scattered light from a single pentamer was collected using dark-field spectroscopy. The Fano resonance was identified as a narrow and asymmetric dip imposed on a broad spectral profile as shown in Fig. 3b. This effect resulted from the destructive interference between a superradiant bright mode, in which the dipolar plasmons of all the AuNPs oscillated in phase, and a subradiant dark mode, in which the small nanosphere capacitively coupled with two of the nanoshells (see Fig. 3b). In the quasistatic, nonretarded limit, the dark mode should possess nearly no net dipole moment and should be not easily excited by light. In the retarded limit, the bright and dark modes became supperadiant and subradiant, respectively. In other words, the coupling mediated by the plasmonic near-fields led to the interaction between the sub- and superradiant modes, introducing a Fano resonance in the superradiant continuum at the frequency of the subradiant mode. The Fano resonance depth is strongly dependent on the constituent particle sizes, shapes, interparticle spacing, the number of particles, as well as the cluster symmetry as demonstrated by a vast of follow-up experiments, mainly carried out using top-down nanotechniques. [99,100] [101] [102]

### 3.3 Plasmonic Magnetic Resonances

Plasmonic magnetic resonances have attracted a lot of attention due to the enthusiastic pursuits of artificial negative refractive index materials, which hold the promise for many fascinating phenomena including invisibility cloaking, super lensing, and negative refraction. [103,104] [105] [106] In general, simultaneous negative permittivity and negative permeability are required to occur in the same frequency region to effectively achieve a negative refractive index. While metals naturally possess negative permittivity at visible frequencies, negative permeability has to be created through an artificial magnetic response. Nevertheless, structural designs such as split-ring resonators, which are used to achieve negative permeability at lower frequencies, [107,108] cannot be easily transferred due to the so-called saturation effect of the magnetic response at visible frequencies. [109] To overcome this problem, Engheta *et al*. came up with an elegant design, in which metal nanoparticles are brought together in a ring geometry. [110] [111] Such a structure can support a circulating displacement current due to plasmon excitations, which can lead to the occurrence of magnetic resonances at high frequencies.

Fig. 4a shows the formation of the magnetic resonances in DNA-assembled quadrumers demonstrated by Fan *et al*.[98] Four Au nanoshell particles were assembled in a ring geometry. The cross-polarized scattering spectrum of the quadrumer exhibited a clear and narrow peak near 1400 nm. The simulated mode plot at this resonance revealed a circulating current around the ring of the nanoshells, which identified the excitation of the magnetic resonance. Later, Roller *et al*. utilized DNA origami to assemble small nanoparticle rings, which exhibited magnetic resonances at visible frequencies as shown in Fig. 4b.[87] A DNA origami structure composed of 14 parallel arranged DNA helices of 200 nm length was designed to curve into a ring by insertion and deletion of bases at selected sites.[75] The ring had a diameter of 62 nm and a cross-section of ~10 nm. Four AuNPs (40 nm) covered with ssDNA were then assembled on one origami ring through DNA hybridization. The individual plasmonic nanostructures immobilized on a glass substrate were characterized by dark-field spectroscopy. It was demonstrated that the magnetic resonance appeared only in four-particle rings with broken symmetry, which enabled a coupling channel between the magnetic mode and the far-field photons propagation perpendicular to the substrate. The simulated charge and magnetic field distributions confirmed the formation of circulating currents and a magnetic dipole in the particle ring structure, respectively, indicating the excitation of the magnetic resonance as shown in Fig. 4b.

### 3.4 Plasmonic Chirality

A geometrical object is chiral, if its mirror image is non-superimposable with itself. Just like our two hands, which are non-superimposable to each other, many molecules can exist in left-handed (LH) and right-handed (RH) forms, the so-called enantiomers. Chirality is crucially important in pharmaceutics, as the biological activities of enantiomer molecules can be distinctly opposite. One form may be helpful, whereas the other form may be inactive or even toxic.

The commonly used spectroscopic technique to study chiral molecules is circular dichroism (CD), which is defined as the difference in extinction of LH and RH circularly polarized light. Natural chiral molecules such as proteins and DNA exhibit weak CD mostly in the UV spectral range. Govorov *et al*. theoretically proposed that it is possible to build plasmonic analogs of chiral molecules, which show much stronger CD than their molecular counterparts and possess large spectral tunability. [112]

One of the first attempts toward DNA-assembled plasmonic chiral nanostructures was accomplished by Mastroianni *et al*. AuNP pyramids linked by DNA were carefully constructed as shown in Fig. 5a. [54] In an individual structure, each DNA strand was designed with a unique sequence, which allowed for positioning of a single AuNP at one of the pyramidal tips. Four AuNPs of different diameters (5, 10, 15, 20 nm) were arranged in one structure. The enantiomeric version was obtained by switching the positions of two of the AuNPs through conjugation with the opposite strands. Unfortunately, CD was not observed from these plasmonic assemblies mainly due to negligible coupling between the AuNPs of substantially different sizes and comparatively large inter-particle distances. In addition, the non-rigidity of the assemblies led to varying configurations over different nanostructures and therefore vanishing CD.

The power of structural DNA nanotechnology for constructing rigid plasmonic architectures was first demonstrated by Kuzyk *et al*. [20] The authors designed two 100 nm long DNA origami bundles of 24 DNA helices as templates to organize 10 nm AuNPs in LH and RH helical forms, respectively (see Fig. 5b). Capture strands were extended from nine binding sites on each of the two templates. Through DNA hybridization, AuNPs functionalized with complementary DNA were assembled around the DNA origami bundles in a staircase fashion, forming LH and RH AuNP helices, respectively. Due to the near-field interactions between the AuNPs in a 3D chiral order, these plasmonic assemblies gave rise to a bisignate CD response, which resembled that of chiral molecules, however, was orders of magnitude stronger and became apparent at visible frequencies. As predicted by theory, the LH and RH AuNP helices exhibited mirrored CD spectra. The authors also demonstrated the tunability of the CD response by Ag enhancement of the AuNP helices. In an approach by Shen *et al.*, a rectangular DNA origami sheet dressed with thirteen AuNPs forming two parallel lines was rolled up into a tubular shape as shown in Fig. 5c. [19] As the repulsion between the particles favored the closure of the sheet toward one direction, AuNP helices of defined handedness were formed and shown to exhibit a measurable CD response in the visible.

As of today, a great variety of DNA origami-templated chiral plasmonic architectures have been realized. For example, Shen *et al.* also demonstrated DNA-assembled plasmonic tetramers (see Fig. 5d). [91] Each tetramer contained four identical 20 nm AuNPs assembled on a rectangular DNA origami template. In contrast to the aforementioned work of Mastroianni *et al.*, the four AuNPs in the plasmonic tetramer could strongly couple due to their identical size. Also, the rigidity of the origami template and the utilization of multiple capture strands per AuNP contributed to the homogeneity of the structural configurations, allowing for experimental observation of distinct CD from the LH and RH plasmonic tetramers.

DNA origami is ideally suited to construct chiral systems with increasing complexities, for example, a plasmonic toroidal structure (see Fig. 5e), [90] which is extremely challenging to achieve using top-down nanotechniques. Four curved origami bundles were linked in a ring geometry. Twenty four AuNPs (13 nm) functionalized with complementary DNA strands compared to the capture strands extended from the origami ring were assembled in a helical fashion to form a LH or RH plasmonic toroidal structure. The assembled plasmonic toroids with designated handedness exhibited pronounced optical activity in the visible spectral range. In addition, the authors demonstrated that given the unique circular symmetry, distinct chiroptical response along the axial orientation of the toroids could be revealed by simple spin-coating of the structures on substrates. Such orientation self-alignment neither required surface functionalization nor introduced birefringence effects, which often arise from symmetry breaking.

The dominating role of Au nanospheres as key player for building DNA-assembled plasmonic nanostructures was terminated after the successful demonstration of well-controlled gold nanorod (AuNR) assemblies on origami by Pal *et al*. [114] Due to their strong optical response and anisotropic nature, AuNRs are excellent candidates for the realization of advanced plasmonic architectures with distinct and tailored optical functionalities. Soon after the work of Pal *et al*., Lan *et al*. realized crossed AuNR (11 nm × 37 nm) dimers templated by bifacial DNA origami as shown in Fig. 3f. [92,115] Importantly, the 3D spatial configuration of the dimer could be precisely tuned by rationally shifting the AuNR positions on origami. The two crossed AuNRs constituted a 3D plasmonic chiral object, which generated a theme of handedness when interacting with LH and RH circularly polarized light, giving rise to strong CD. As a matter of fact, the CD response from such AuNR dimers can be well described by the Born-Kuhn model, which was already used to interpret the phenomenon of optical activity in natural chiral molecules. A coupled AuNR dimer mimics the geometry of a molecule containing two chromophores, a situation commonly described as exciton-coupling in organic chemistry. Taking a step further, Lan *et al*. also created AuNR helices, which contained twisted AuNRs directed by DNA origami as shown in Fig. 5g [113]. Capture strands in an 'X' pattern were arranged on the two sides of a 2D origami template. AuNRs functionalized with complementary DNA sequences were positioned on the origami and subsequently led to AuNR helices with the origami intercalated between neighboring AuNRs. LH and RH AuNR helices were conveniently accomplished by tuning the mirrored-symmetric 'X' patterns of the capture strands on the origami. These DNA-assembled plasmonic chiral nanostructures shed light on many interesting applications including tunable chiral fluids, enantiomer sensing, chiral signal amplification, and fluorescence-combined chiral spectroscopy.

## 3.5 Surface-Enhanced Raman Spectroscopy

Surface-enhanced Raman spectroscopy (SERS) is a surface sensitive technique that utilizes metal surfaces or nanostructures to enhance Raman scattering of adsorbed molecules. [116,117] The ability to analyze specific chemical fingerprints of minute amounts of molecules on the nanoscale makes SERS extremely useful for applications in environmental analysis, pharmaceuticals, material sciences, drug and explosive detection, as well as food quality analysis. [118] However, SERS substrates often encounter poor reproducibility and performance in that the Raman scattering cross-sections of the probed molecules can be orders of magnitude smaller than fluorescence cross-sections. [119] [120] Fortunately, SERS signals can be greatly enhanced by placing Raman-active molecules in plasmonic hot spots, where intense electromagnetic fields are highly localized. [121] [122] Such SERS intensity is proportional to the fourth power of the localized electric field with a possible enhancement factor beyond $10^{10}$. [123] In general, the electromagnetic fields in plasmonic hot spots are enhanced as the nanogap size decreases, especially down to the 1 nm region. [27,35,93] Therefore, over the last decades continuous efforts have been exerted to synthetize various plasmonic nanostructures possessing narrow gaps for achieving strong and reliable SERS signals. [28] [124]

The successful synthesis of DNA-assembled Au core/Ag shell heterodimers was reported by Lim *et al*. [27] These particles could be used to detect Raman signals from single dye molecules placed in the dimer gaps as shown in Fig. 6a. AuNP (20 and 30 nm) heterodimers were first synthesized by means of DNA hybridization. A single Cy3 dye molecule was then conjugated between the two DNA-tethered AuNPs. Ag shell growth on the AuNPs was carefully controlled to generate gap-engineerable core-shell dimers. Such DNA-assembled heterodimers manifested remarkable single-molecule sensitivity with high structural reproducibility. When modified by other biomolecules such as proteins, these heterodimers could further serve as both *in vitro* and *in vivo* bio-labelling probes with ultrahigh sensitivity, quantification potentials, and multiplexing capability. Furthermore, another synthetic strategy has been developed by the same group for preparing Au-Ag nanosnowman structures with crevice nanogaps as shown in Fig. 6b. 40 nm AuNPs as seeds were modified with thiolated-DNA. [125] Ag precursors and other reagents were then added for the asymmetric growth of AgNPs on the DNA-AuNP surfaces. Such nanosnowman particles could generate strong SERS signals in that the narrow conductive junctions allowed for plasmonically enhanced electromagnetic fields and Raman-active molecules could be locally positioned by DNA in the hotspots. Notably, the synthetic yield of the nanosnowman particles was as high as 95%, greatly suitable for SERS-based chemical and biological sensing.

It is worth mentioning that Lim *et al*. also synthesized a new class of plasmonic nanostructures termed Au nanobridged nanogap particles (Au-NNPs) as shown in Fig. 6c. [27] The Au-NNP had a hollow, 1 nm interior gap between a Au core and a Au shell. The process involved modification of AuNPs with DNA possessing Raman dyes and thiols. The thickness of a single strand of DNA is ~ 1nm, matching the interior gap size. The Au nanobridges that connected the core and the shell were mainly governed by the sequence of the thiolated DNA bases and the DNA grafting density. The precise and quantitative positioning of Raman dyes inside the narrow gap of the Au-NNP led to strong, highly reproducible SERS signals. Later, Kang *et al*. successfully applied such Au-NNPs, responding to near-infrared excitation (785 nm), in combination with high-speed confocal Raman microscopy for live cell Raman imaging. [126] These particles allowed for monitoring spatial particle distributions within their targeted sites such as cytoplasm, mitochondria, or nucleus without inducing significant damages on cells as well as for detecting rapidly changing cell morphologies induced by addition of highly toxic potassium cyanide to cells.

As DNA origami represents a reliable fabrication platform to create probes of strongly coupled plasmonic nanostructures with well-defined geometries, several research groups have already implemented DNA origami-templated AuNP assemblies for SERS. [29] [31] As an extra benefit, the rigidity of the origami surface prevents samples from degradation and ensures stability during SERS measurements. The first Raman signals originating from a DNA origami platform were measured and reported by Prinz *et al.*[29] In this work, TAMRA-modified DNA strands have been placed between 5 nm gold nanoparticles that were overgrown with additional Au by electroless deposition. Also, Pilo-Pais *et al.* used Au overgrowth on particles assembled in a tetramer on a DNA origami sheet. Subsequently, the structures were incubated with the Raman dye (4-aminobenzenethiol), which covalently attached to the Au surfaces and gave rise to clear signals. [31] In particular, Thacker *et al*. demonstrated the assembly of 40 nm AuNP dimers with sub-5 nm gaps on a DNA origami platform as shown in Fig. 6d. [30] This design allowed for a strong coupling between the two AuNPs. Individual dimers were immobilized on a Au-coated silicon wafer and then briefly incubated in a Rhodamine 6G solution to form a monolayer on the dimer structures. The observed enhancement was highly sensitive to the polarization of the laser with respect to the dimer axis. Five molecules were estimated to contribute to the observed SERS signal. The calculated surface enhancement factors were between five and seven orders of magnitude. Almost at the same time, Kühler *et al*. also demonstrated AuNP dimers assembled by DNA origami for SERS applications (see Fig. 4e). [32] Their AuNP (40 nm) dimers were linked together by a three-layered DNA origami block at a separation distance of 6 nm to achieve plasmonic coupling and the formation of a plasmonic hot spot. Different from the work of Thacker *et al*., the authors located Raman-active molecules of interest by selectively incorporating SYBR Gold only in the DNA structure and thus in the hot spot region between the AuNPs. SYBR Gold is a minor groove-binding fluorescent nucleic acid stain that has a high affinity to double-stranded DNA. The measured SERS signal was estimated to originate from ~ 25 dye molecules localized in the hot spot region.

In order to push the sensing limit towards single molecule resolution, further reduction of the interparticle distance between two AuNPs to achieve higher field enhancements was carried out by Simoncelli *et al*. as shown in Fig. 6f.[93] The authors used optothermal-induced shrinking of a DNA origami template to control the gap sizes between two 40 nm AuNPs in a range from 1 to 2 nm. Before laser irradiation, the Raman spectrum of a single Cy3.5 molecule placed in the center of the hotspot did not show any visible peaks. Upon irradiation with a continuous-wave 612 nm laser (60 kW/cm$^2$) for 10 s, a red shift of the scattering peak was observed corresponding to reduction of the interparticle gap. Signature peaks for the single Cy3.5 dye were observable in the Raman spectrum. The gap sizes were inferred using Mie theory calculations according to the measured scattering spectra before and after laser irradiation. The numerical simulations showed that the gap size was reduced from ~ 3.3 to 1.9 nm. The intensities of the Raman peaks were more pronounced after repeating the heating step for a second time. The scattering peak of the Au dimer was further red-shifted by about 30 nm. The gap was estimated to be further reduced from ~1.9 to 1.3 nm. This work, together with the work by Prinz *et al.*, who used Ag overgrowth on DNA origami-assembled AuNPs to obtain SERS signals from single molecules, [33] outlines the path towards achieving reliable plasmonic platforms for single-molecule SERS detection.

### 3.6 Surface-Enhanced Fluorescence Spectroscopy

Quantifying the interplay between single emitters and plasmonic nanostructures provides important insight into the underlying physics of light-matter interaction on the nanoscale. Among different interaction mechanisms, surface-enhanced fluorescence holds tremendous potential for detection of single molecules, biosensing applications, and nanoscale light control. Plasmonic nanostructures, which on the one hand may efficiently alter the local electric fields in hot spots, can on the other hand directly affect the excitation rate of a fluorophore as well as influence its emission properties such as quantum yield and angular emission pattern. [127] [128,129] [36,130] All these enhancement effects are highly dependent on the plasmonic nanostructure's size, geometry, material, and relative position to the fluorphore. In the past, significant efforts have been made to place single molecules into hot spots of plasmonic nanostructures. Unfortunately, most of the experimental schemes failed to provide nanoscale positioning accuracy. In contrast, DNA assembly intrinsically offers nanoscale docking sites that are fully addressable, opening a unique pathway to study light-matter interaction with unprecedented control.

Pioneering work toward this direction was carried out by Acuna *et al.*, who used a tower-shaped DNA origami structure as shown in Fig. 7a. [36] This origami platform allowed for precise assembly of a AuNP (100 nm) dimer with a defined interparticle distance and simultaneously provided a docking site for locating fluorescent dyes in the plasmonic hot spot. The 12-helix bundle tower had a height of 220 nm and a diameter of 15 nm. Three additional 6-helix bundles widened the base of the origami tower to ensure its stable immobilization on a cover slip through biotin-streptavidin conjugation. [77] Two DNA-functionalized AuNPs were hybridized to three capture strands each extending from the sides of the origami tower at pre-designed sites. A dye-labeled strand (ATTO647N) was incorporated in the origami at the center of the plasmonic hot spot during assembly such that the dye was aligned with respect to the electric field polarization. The authors observed a maximum of 117-fold fluorescence enhancement for the dye molecule positioned in the 23 nm gap between two 100 nm AuNPs. Later, the same group also reported optimized origami platforms (see Fig. 7b) by improving the structural robustness, reducing the interparticle distance, *etc*. Fluorescence enhancement of more than 5000-fold was achieved. [39] Evidently, such DNA origami platforms outperformed top-down lithographic ones. Also, the authors avoided the problem of bleaching the dye molecules by using short, dye-modified oligonucleotides that only transiently bound to an anchor sequence protruding from the position of interest on the origami. [131]

## 4. DNA-ASSEMBLED DYNAMIC PLASMONIC NANOSTRUCTURES

DNA-assembled dynamic plasmonic nanostructures refer to DNA-assembled plasmonic nanostructures, which not only yield well-defined optical properties but also possess desired structural reconfigurability. Such plasmonic devices hold great promise for applications in adaptable nanophotonic circuitry, artificial nanomachinery, as well as optical sensing of molecular binding and interaction activities. DNA structures generally allow for various ways to control their dynamic behavior. Basic schemes for structural reconfigurations of DNA structures can utilize DNA hybridization/dehybridization as well as pH and ion concentration stimuli.

Probably, the most versatile and thus widespread approach is the so-called "toehold-mediated strand displacement" scheme as shown in Figure 8. In this process, a freshly added DNA sequence binds to a short extension, *i.e.*, the toehold, on the DNA structure and replaces another sequence through branch migration. [12] More intriguing schemes include enzymatic cutting of pivotal connector strands or structural adaptions of aptamers to the presence of target molecules. [132] [133] Also, photoresponsive molecules can be employed through incorporation with DNA to alter the strand's hybridization kinetics in response to light stimuli. [134] [135]

## 4.1 Plasmon Rulers

Plasmon rulers are used to monitor distance changes with nanometer precision. The underlying mechanism can be explained by the plasmon hybridization theory: when two metal nanoparticles are placed into proximity, their plasmons couple to each other. [80] [81] Depending on the interparticle separation, the resonance wavelength of the coupled system shifts sensitively. Importantly, plasmon rulers represent significant advantages over molecular rulers based on Förster Resonance Energy Transfer (FRET). [136] [137] [138] FRET often suffers from low and fluctuating signal intensities as well as photobleaching. In contrast, plasmon rulers neither blink nor bleach, offering unlimited observation time. In addition, FRET can only sense distance changes within 10 nm and the signal almost drops in a step-wise fashion, which often hampers quantitative interpretations. The range of accessible distances for plasmon rulers on the other hand can be up to several tens of nanometers, while the shifts of the plasmon resonances can occur more or less continuously. [3]

The first plasmon rulers were created by Sönnichsen *et al.* using DNA-assembled 40 nm AuNP or AgNP dimers as shown in Fig. 9a. [22] The authors utilized surface-immobilized particles as anchors for ssDNA-functionalized particles. The ssDNA molecules were bound *via* thiol groups to the free AuNPs and had biotin at their other ends, allowing them to bind to the streptavidin-coated anchor particles. Upon binding, the coupled scattering centers suddenly changed color under dark-field illumination. The authors further demonstrated that DNA-linked AuNP dimers could be used to report interparticle distance changes in real time. They detected the hybridization of a complementary DNA oligonucleotide to the flexible ssDNA linker within a AuNP dimer through a significant blue-shift of the resonance position, which resulted from the stiff doubled-stranded DNA pushing apart the AuNPs. As a result, the dynamics of DNA hybridization could be recorded *in situ* by monitoring the spectrum of a single AuNP dimer continuously.

After this pioneering work, a variety of plasmon rulers designed for different applications have been created. For example, Reinhard *et al*. applied a plasmon ruler to measure the dynamical biophysical processes during the cleavage of DNA by the restriction enzyme EcoRV on the single molecule level (see Fig. 9b). [139] A plasmon ruler composed of two linked 40 nm AuNPs was immobilized on the surface of a glass flow chamber *via* a biotin-Neutravidin bond between the biotin-functionalized particle and surface. The color and intensity of the plasmon ruler were monitored in real time using dark field microscopy. Upon addition of the enzyme, some plasmon rulers exhibited sudden intensity drops due to enzymatic DNA cleavage. Intriguingly, the authors could also demonstrate that the observed intensity increase before the cut was due to DNA bending by the EcoRV restriction enzyme. These remarkable results were only possible to obtain, given the unlimited lifetime, high temporal resolution, and high signal/noise ration of the plasmon rulers, which substantiate a unique tool for studying conformation changes of molecular events and subsequent dynamic activities at the single molecule level.

Plasmon rulers with more complex geometries have also been attempted. For example, Sebba *et al.* demonstrated reconfigurable core-satellite assemblies as molecularly-driven plasmonic switches (see Fig. 9c). [140] A 50 nm Au core particle and 13 nm Au satellite particles were linked together using DNA, which contained a hairpin-shaped secondary structure for modulating the duplex length. Reconfiguration of the core-satellite assemblies from the hairpin state to the extended state through strand displacement reactions led to a significant decrease of the scattered intensity and a restoration of the plasmon resonance position to that of isolated core particles. Another interesting work was demonstrated by Morimura *et al.*, who utilized AuNP dimers to report DNA conformation changes induced by transcription factor binding (see Fig. 9d). [141] The AuNP dimers bound by SOX2 shifted the plasmon resonance position to a longer wavelength, indicating bending of the DNA duplex induced by SOX2 binding. When the SOX2 formed a ternary complex with PAX6 on DC5, a further resonance red-shift was observed, implying additional bending in the DC5 sequence.

Other than utilizing resonance positions to correlate nanoscale distance changes, AuNPs assembled with DNA in a chain geometry can particularly work as an energy transfer platform for guiding electromagnetic energy below the optical diffraction limit. Recently, Vogele *et al.* have developed a switchable plasmonic waveguide composed of AuNPs assembled on a DNA origami structure that allowed for simple spectroscopic excitation and readout (see Fig. 9e). [142] Multiple AuNPs were positioned along a line on a DNA origami structure. The central AuNP was functionalized with thermoresponsive elastin-like peptides (ELPs). ELPs are synthetic peptides that exhibit a fully reversible hydrophobic collapse upon a certain transition temperature. One outer AuNP was functionalized with fluorescein (FAM) and it worked as the excitation port. Near-field coupling among the adjacent AuNPs enabled the electromagnetic energy transfer along the particle chain. Evaluation of the waveguide performance was achieved by detecting the emission of an acceptor dye (Atto 532), which was located on the surface of the other outer AuNP. Below the transition temperature, the coupling between the AuNPs was weak as the highly solvated and thus swollen ELPs pushed the central AuNP out of line. With a temperature increase above the transition temperature, the ELPs were transformed into the collapsed state leading to shortening of the particle spacing. The resulting stronger plasmonic coupling among the AuNPs facilitated the energy transfer, reflected by a slight increase of the fluorescence signal. Owing to the reversibility of the ELP swelling and collapsing, the fluorescence signal of the waveguide could be repeatedly switched between two states by temperature cycling.

**4.2 Dynamic Manipulation of Plasmonic Chirality**

Controlling molecular chirality is of great importance in stereochemistry. Chirality of natural molecules can be manipulated by reconfiguring molecular structures through light, electric field, and thermal stimuli. However, such chirality regulation lacks efficiency as the CD response of natural molecules is very weak. In contrast, plasmonic chiral nanostructures assembled by DNA allow for dynamic manipulation of chirality and reversible switching of strong CD responses.

Kuzyk *et al*. reported the first reconfigurable 3D plasmonic nanostructures, which executed DNA-regulated conformational changes on the nanoscale (see Fig. 10a). [89] Two AuNRs were hosted on a reconfigurable DNA origami template, which consisted of two linked 14-helix bundles (80 nm × 16 nm × 8 nm). Twelve anchor strands were extended from each origami bundle for robust positioning of two AuNRs (40 nm × 10 nm, one on each bundle) functionalized with complementary DNA. Two DNA locks that were utilized to manipulate the configuration of the plasmonic nanostructure were extended from the sides of the DNA origami bundles. Each lock contained two DNA arms, which could be opened or closed through toehold-mediated strand displacement reactions upon addition of corresponding fuel strands. As a result, the configuration of the plasmonic nanostructure could be switched between LH, RH, and relaxed states. Importantly, the 3D plasmonic nanostructures worked as optical reporters, which transduced their configuration changes *in situ* into CD changes at visible frequencies.

Later, Kuzyk *et al*. also demonstrated a light-driven 3D plasmonic nanosystem that could translate molecular motion of photon-switchable azobenzene into reversible chiroptical function (see Fig. 10b). [143] Introducing an azobenzene-modified DNA segment into a reconfigurable plasmonic DNA origami nanostructure enabled photoactive function. The photoresponsive segment comprised two DNA branches, which were extended from the two origami bundles, respectively. One branch possessed a 20-base-pair-long double-stranded segment that was linked by a disulfide bond to a segment comprising azobenzene-modified nucleotides (Azo-ODN 1). The other branch contained the pseudocomplementary strand Azo-ODN 2. Upon UV light illumination, the azobenzene molecules in both Azo-ODNs were converted to the *cis*-form, resulting in dehybridization of the Azo-ODN duplex and opening of the origami template. In contrast, upon illumination with visible light, the azobenzene molecules were converted to the *trans*-form and the Azo-ODNs could hybridize again thus closing the origami template. The different conformation states could cyclically be induced and be read-out by probing light in real time. This system thus amplified the sub-nanometer conformation changes of azobenzene through large-scale changes of the host nanostructure and consequently translated the light-triggered molecular motion of azobenzene into strong and reversible plasmonic CD responses.

Dynamic CD switching has also been achieved by manipulating the orientations of plasmonic chiral nanostructures relative to the incident light direction. As shown in Fig. 10c, Schreiber *et al*.

mounted DNA origami-assembled LH AuNP helices on a glass substrate through biotin-neutravidin binding. [88] CD measurements were then performed by placing the substrate perpendicular to the light beam inside a cuvette that was initially filled with buffer. The AuNP helices stood upright on the substrate and hence were aligned parallel to the probing light beam. Instead of a peak-dip CD of the LH helices isotropically dispersed in solution, an inverted dip-peak spectrum with a dominant peak was observed, which corresponded to the axial component of the averaged CD signal ($CD_z$). In the next step, the sample was dried with nitrogen and placed again into the optical path. This resulted in the alignment of the AuNP helices perpendicular to the light beam and the recorded CD spectrum showed a peak-dip signal with a dominant dip, *i.e.*, $CD_{xy}$. Reversible switching between two alignment states and consequently switching between the signals of $CD_{xy}$ and $CD_z$ could be achieved by repeated flushing and drying processes without changing the orientation of the substrate.

The high sensitivity of CD on conformation changes also enables optically tracking of successive movements of plasmonic nanoparticles well below the diffraction limit. Zhou *et al*. demonstrated a dynamic plasmonic system, in which one AuNR could perform stepwise walking directionally and progressively on DNA origami (see Fig. 10d).[144] A walker AuNR and a stator AuNR (an immobilized AuNR) were placed on two opposite surfaces of a rectangular DNA origami platform, forming a chiral geometry. While the stator was immobilized on one surface, the walker rod on the other surface could execute stepwise, directed movements. Along the track that was laid out on the origami surface, six parallel rows of footholds were utilized to establish five walking stations, evenly separated by 7 nm. The stepwise walking was powered by DNA hybridization and triggered by the addition of the respective blocking and removal strands. The dynamic walking process could be monitored *in situ* using a time-scan function of the CD spectrometer at a fixed wavelength. In the experiments, the plasmonic walker not only functioned as a walking element to carry out mechanical motion but also as an optical reporter, which could deliver its own translocation information through optical spectroscopy in real time. Later on, the same group demonstrated the implementation of two plasmonic walkers that could walk both individually and collectively on the same origami track (see Fig. 10e). [145] A sensitive plasmonic coupling scheme was introduced for in situ optically monitoring the dynamic walking of the two walkers with steps well below the diffraction limit. The two walkers and a stator were assembled on the two opposite surfaces of an origami template. The two walkers optically interact with the stator simultaneously. In particular, each walker and the stator constituted a chiral geometry. The chiroptical response of the entire system was jointly determined by the positions of both walkers relative to the stator. This rendered optical discrimination of the walking directions and the individual steps of the two walkers possible. In addition, the walker number and the optical response of the system could be correlated.

Very recently, selective control of different plasmonic nanostructure species coexisting within one ensemble has been demonstrated by Kuzyk *et al*. as shown in Fig. 10f. [146] Reconfigurable chiral plasmonic nanostructures were assembled using DNA origami in LH and RH states, respectively. In these structures, pH-sensitive DNA locks worked as active sites to trigger structural regulation over a wide pH range. Such locks exploited triplex DNA secondary structures that display pH-dependent behavior due to the presence of specific protonation sites. The selective reconfiguration was enabled by modulating the relative contents of TAT/CGC triplets in the DNA locks. To demonstrate the unprecedented enantioselectivity, the authors mixed LH and RH structures in equimolar amounts to form a quasi-racemic solution. Such plasmonic quasi-enantiomers, *that is*, enantiomers as plasmonic objects, but functionalized with DNA locks containing different TAT contents could be selectively activated upon pH tuning, elucidating an innovative approach to arbitrarily modulate chiroptical response at wish.

## 4.3 Dynamic Plasmonic Nanostructures for Sensing Applications

Due to the high sensitivity of plasmonic hot spots, the engineerable dynamic properties, and the unique biological specificity, DNA-assembled dynamic plasmonic architectures open the route for a new generation of optical sensors, which are vital for fundamental cellular studies and clinical diagnostics. Due to their ease of synthesis and sensitive read-out mechanisms, DNA-assembled plasmonic architectures are particularly attractive for analyte binding and molecular sensing.

Chen *et al*. demonstrated that target-induced actuation of DNA-assembled AuNP dimers could be an effective mechanism for DNA molecule sensing based on the plasmon-hybridization scheme (see Fig. 11a).[147] DNA-assembled plasmonic structures have also been used to sense gas molecules such as hydrogen. Li *et al*. reported DNA-assembled bimetallic plasmonic nanostructures composed of AuNRs and palladium nanoparticles (see Fig. 11b).[148] The DNA strands served both as linkers and seeding sites for the growth of the palladium nanoparticles and facilitated reliable positioning of palladium satellites around a AuNR with an ultrashort spacing of several nm. Upon hydrogen absorption, palladium underwent a phase transition to form palladium hydride, which then altered the neighboring environment of the centered AuNR. Using dark-field spectroscopy, these plasmonic superstructures were able to efficiently detect dynamic absorption and desorption of hydrogen on the single structure level.

More intriguingly, Xu *et al*. reported DNA-assembled Ag pyramids for simultaneous and ultrasensitive detection of multiple disease biomarkers such as prostate specific antigen, thrombin, and mucin-1 as shown in Fig. 11c. [149] Here a DNA frame holding aptamers specific to disease biomarkers was used to drive the self-assembly of AgNP pyramids dressed with Raman-active molecules. Upon binding of the designated disease markers to the aptamers, the 3D configuration of the pyramids could be altered. This process resulted in a shorter interparticle separation, which in turn caused Raman signal enhancement. Similarly, Kotov *et al*. produced DNA-assembled plasmonic superstructures composed of AuNRs and spherical AuNPs, which contained a varying number of plasmonic gaps (see Fig. 11d). [150] The authors applied these superstructures in live cells for SERS-based *in situ* monitoring of intracellular metabolism. Incubation of these superstructures with Hela cells indicated sufficient field enhancement to detect structural lipids of mitochondria and potentially small metabolites.

Chiral plasmonic nanostructures are a newly developed member in the family of optical sensors. One of the unique characters afforded by chiral plasmonic nanostructures is their high sensitivity on minute configuration changes. Also, their 3D nature renders possible sensing of analyte binding and other biochemical-related activities with extra degrees of freedom. For example, Yan *et al*. demonstrated a pyramidal sensor platform with reversible chiroptical signals for DNA detection (see Fig. 11e). [151] In the presence of target DNA, two types of nanoparticle pyramids underwent dynamic reconfiguration or a dissociation process, which functioned as an off-on or on-off switch towards CD signal changes. Target DNA could be detected at the attomolar level. Later, the same group also reported the realization of Au-upconversion nanoparticle pyramids to detect intracellular microRNA in real time. [152]

## 5. CONCLUSION, OUTLOOK, AND PERSPECTIVE

The benefits of self-assembled plasmonic systems are manifold and evident: literally nanometer-precise positioning of optical elements is paired with massive parallel assembly, which allows for the production of large quantities of custom-made meta-molecules. This enables plasmonic materials to be structured in all three dimensions on the nanoscale and can thus perform in the visible frequency domain in the realm of possibility. Components of different types – *e.g*. AuNPs, AgNPs, quantum dots, organic dye molecules, analytes, *etc.* – are readily combinable and can be assembled with high yields. Furthermore, the crystalline quality of colloidal particles usually surpasses that of top-down deposited materials, which results in superior plasmonic performance of the devices. There is also a great potential for further exploring the material aspect beyond Au, Ag, *etc*. Plasmonic nanoparticles suffer from absorption losses of metals at optical frequencies. A new promising route to overcome this hurdle has emerged from the discovery of electric and magnetic Mie-type resonances in high-permittivity all-dielectric nanoparticles made of Si, SiC, $TiO_2$, *etc (cite Nature communications 4, 1527 (2013) and Science, 354, 846 (2016))*. Therefore, research efforts on developing feasible and reliable functionalization protocols for assembly of such dielectric nanoparticles on DNA origami will be extremely valuable, as this would lead to novel low-loss nanophotonic devices at optical frequencies. Finally, the development of dynamic plasmonic systems is just beginning and here DNA-based assembly opens ample opportunities for switchable systems that operate autonomously and react to chemical or optical cues. One of the futuristic directions using dynamic plasmonic systems could be sensing of molecular interactions (cite NL 16, 7891 (2016), Science advances 2, e160097 (2016), Science 354, 305 (2016)) on the single molecule or single structure level and thereby readily convert the dynamic interaction behavior into optical signal changes in real time. This is particularly beneficial for the systems, in which large molecules or binding activities from multiple molecular entities are involved, as plasmonic interactions offer a much larger sensing distance range than FRET.

So far, DNA-assembled plasmonic devices have been either operating as individuals in solution or randomly distributed when attached to a surface. Outstanding exceptions are the works of Wallraff, Cha, Gopinath, Rothemund and others, who combined the benefits of DNA self-assembly with top-down methodologies by lithographically rendering surfaces into DNA binding patches surrounded by DNA repelling areas. [153] [154] [155] With this approach the Gopinath *et al.* placed and oriented DNA origami triangles in photonic crystal cavities and beautifully illustrated the achievable control of light-matter interactions by re-painting van Gogh's "The Starry Night" on their nanoscale canvas. [156] In the near future, we expect to see more of such combinations of nanotechnologies but also the combination of hierarchical and large-scale DNA assemblies with plasmonic metamolecules, which could this way be ordered into 2D or 3D arrays. Also, an important objective in the field of DNA nanotechnology is to enlarge the available surface area and mass of origami so that an increasing number of entities can be hosted on origami for achieving more complex and sophisticated devices. There have been attempts to scaling up origami using a biologically derived dsDNA scaffold (cite JACS 131, 9154 (2009), ACS Nano 7, 903 (2013)) or enzymatically producing a ssDNA scaffold (chem communications 48, 6405 (2012)). There is also the so-called 'superorigami' strategy, in which hundred-nanometer sized origami components can be linked together using ssDNA with bridge strands (cite Nano Lett. 2011 Jul 13; 11(7): 2997–3002). The original motivation of Nadrian Seeman to assemble molecules in 3D lattices [9] has driven the field of DNA self-assembly for decades and recent achievements include the formation of rationally designed DNA crystals and large-area lattices build from DNA origami or single-stranded tiles. [157] [158] [159] Linking plasmonic architectures to such DNA lattices will help to establish order in the inherently non-ordered world of self-assembly and open the doors to studies on dynamic metasurfaces and metamaterials operating at visible wavelengths.

**AUTHOR INFORMATION**

**Corresponding Authors**

*na.liu@kip.uni-heidelberg.de; *Tim.Liedl@physik.lmu.de

**Notes**

The authors declare no competing financial interest.

**Biographies**

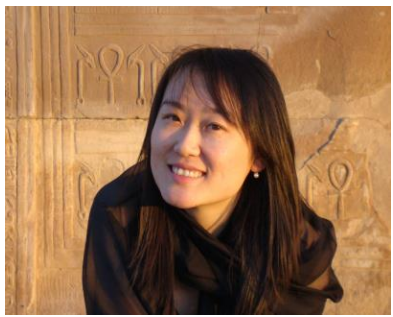

Na Liu is Professor at the Kirchhoff Institute for Physics at the University of Heidelberg. She received her Ph. D in Physics at the University of Stuttgart in 2009, working on 3D complex plasmonics at optical frequencies. In 2010, she worked as postdoctoral fellow at the University of California, Berkeley. From 2011 until 2012, she has worked at Rice University as Texas Instruments visiting professor. At the end of 2012, she obtained a Sofja Kovalevskaja Award from the Alexander von Humboldt Foundation and became an independent group leader at the Max-Planck Institute for Intelligent Systems in Stuttgart. She joined the University of Heidelberg in 2015. The research of Na Liu is multi-disciplinary. She works at the interface between nanoplasmonics, biology, and chemistry. Her group focuses on developing sophisticated and smart plasmonic nanostructures for answering structural biology questions as well as catalytic chemistry questions in local environments.

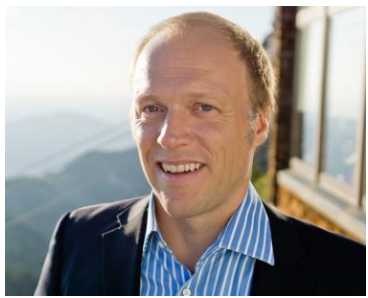

Tim Liedl is Professor for experimental physics at the Ludwig-Maximilians Universität since 2009. During his undergraduate studies he worked on the development of hydrophilic coatings for fluorescent semiconductor nanoparticles. In 2007 he obtained his Ph.D. in physics at the Ludwig-Maximilians Universität studying DNA-based nanodevices and switches which are driven by chemical oscillations. From spring 2007 till summer 2009 he visited Dana-Farber Cancer Institute / Harvard Medical School where he used the DNA-origami method to construct self-assembling two- and three-dimensional structures. The research of Tim Liedl is multi-disciplinary and exploratory positioned at the interface between nanoscience, plasmonics, synthetic biology and cell-biology. Its current focus lies on the application of DNA-based nanostructures in biology and on self-assembled plasmonic materials.

## ACKNOWLEDEMENT

Na Liu acknowledges support from the Sofja Kovalevskaja grant from the Alexander von Humboldt-Foundation, the Marie Curie CIG grant, the Volkswagen grant, and the European Research Council (ERC *Dynamic Nano*) grant. Tim Liedl acknowledges support from the Volkswagen Foundation and the European Research Council through the ERC Starting Grant ORCA (GA N°:336440).

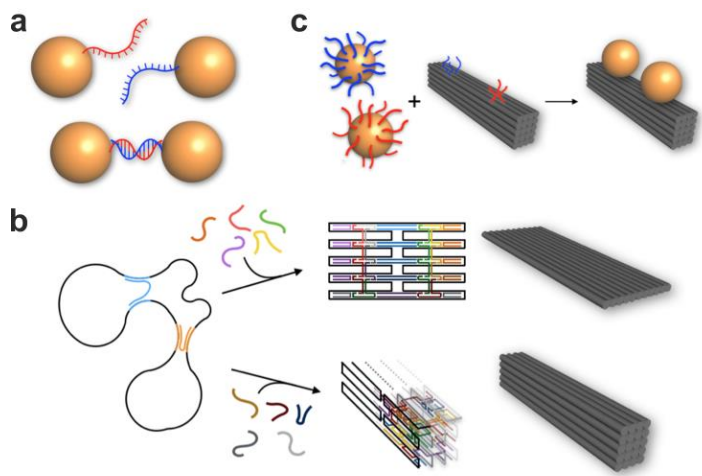

**Figure 1.** Fabrication of DNA-assembled plasmonic nanostructures. a. AuNPs with thiol-modified ssDNA can form a plasmonic dimer through Watson-Crick base-pairing. b. In DNA origami a long scaffold ssDNA strand, typically several thousand nucleotides long, is folded into arbitrary 2D and 3D shapes by hundreds of short staple strands. AuNPs functionalized with single or multiple DNA linkers can be assembled at designated binding sites through hybridization with their complementary DNA strands extending from the DNA origami template to form a wide variety of plasmonic architectures.

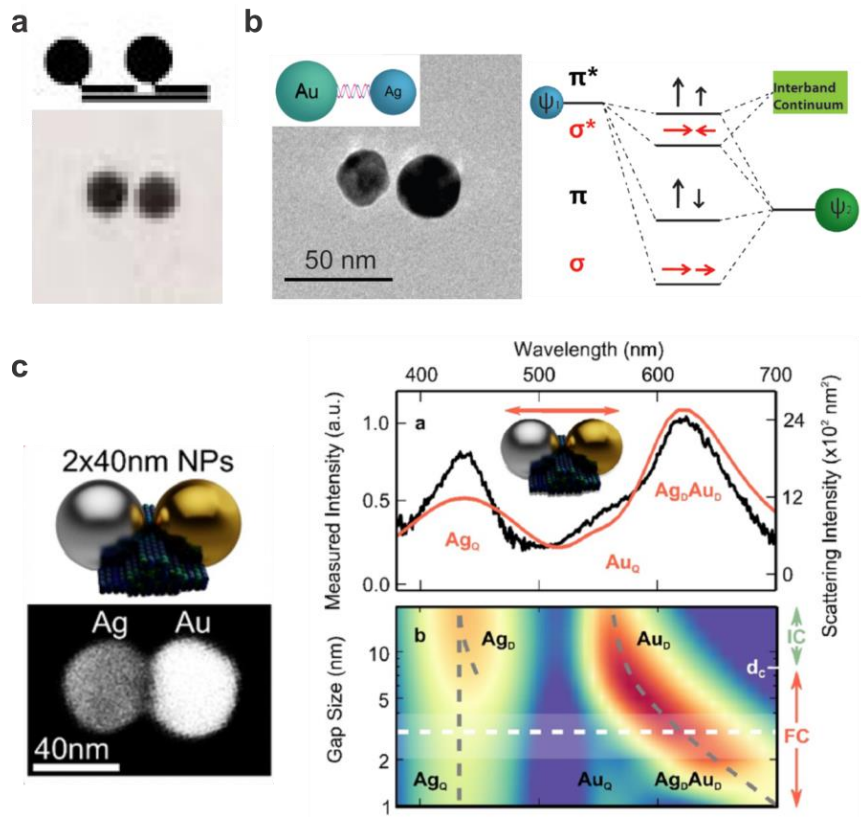

**Figure 2.** Plasmon hybridization in DNA-assembled plasmonic nanostructures. a. AuNP homodimer formation using DNA-monofunctionalized AuNPs.[44] b. Formation of a DNA-assembled Au-Ag heterodimer and the corresponding energy level diagram illustrating the plasmon hybridization.[50] c. Left: Au-Ag heterodimer hosted on a DNA origami template and TEM image of a representative structure. Right: Scattering response of the heterodimers under longitudinal polarization and color map for the scattering response as a function of the gap size.[82]

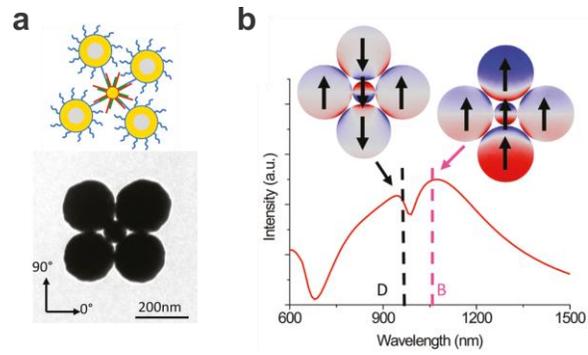

**Figure 3.** DNA-mediated assembly of plasmonic heteropentamers to generate Fano effects. a. Au nanospheres and nanoshells are functionalized with thiolated DNA and incubated together. After drying on a hydrophilic substrate, close-packed pentamers can be formed. A representative TEM image of the structure. b. Extinction spectrum and surface charge plots of the heteropentamer for the dark and bright modes. [98]

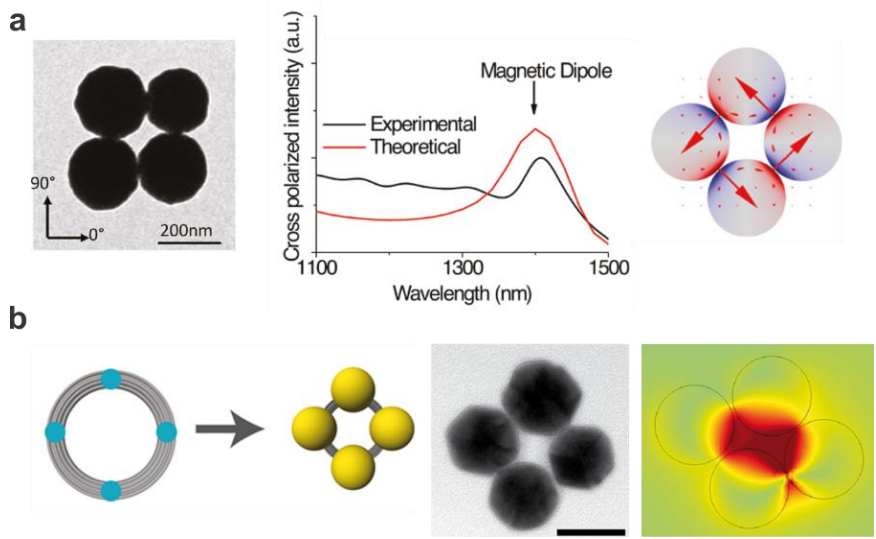

**Figure 4.** Magnetic resonances in DNA-assembled particle clusters. a. Left: TEM image of a DNA-assembled quadrumer. Experimental and theoretical spectra of the DNA-assembled quadrumer, which reveal narrow magnetic dipole peaks near 1400 nm. [98] b. Left: schematic illustration of the DNA origami template for assembly of four AuNPs to form a plasmonic quadrumer and a representative TEM image. Right: simulation of the surface charge distribution for the magnetic resonance. [87]

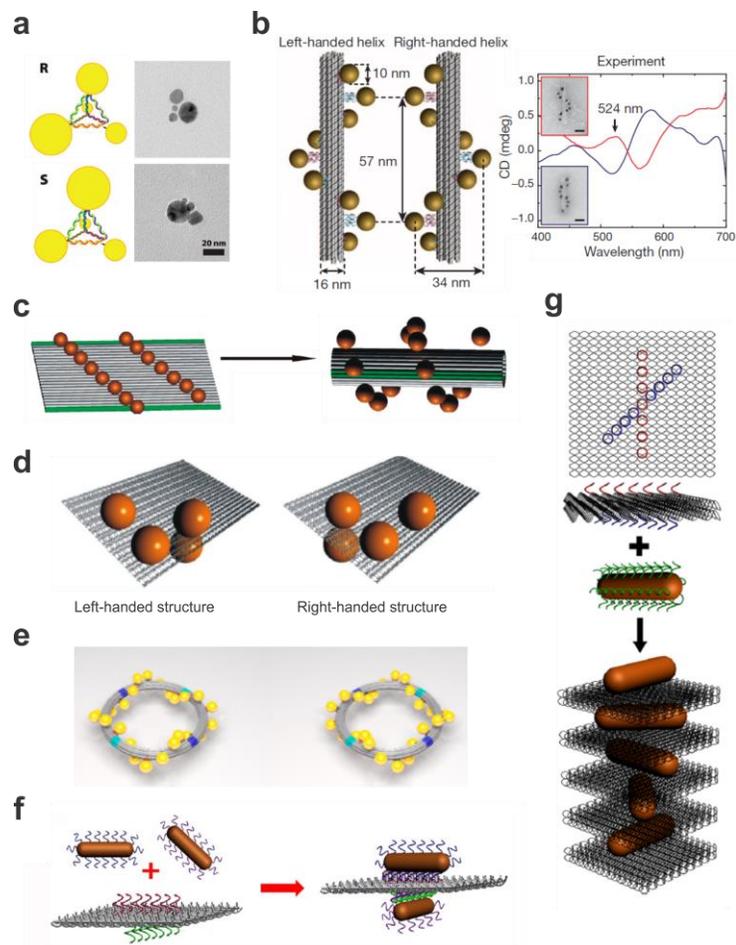

**Figure 5.** DNA-assembled plasmonic chiral nanostructures. a. Schematic and TEM images of the DNA-AuNP pyramids in different handedness. [54] b. LH and RH nanohelices are formed by nine AuNPs that are attached to the surface of DNA origami bundles. AuNPs carry multiple thiol-modified DNA strands, which are complementary to the staple extensions on origami. Experimental CD spectra of the LH and RH helices of nine AuNPs, showing characteristic bisignate signatures in the visible. [20] c. Fifteen AuNPs are assembled on a rectangular origami sheet. Addition of the folding strands leads to rolling and subsequent stapling of the 2D sheet into a hollow tube. As a consequence the AuNPs are arranged into a 3D helix. [19] d. Chiral plasmonic tetramers assembled on DNA origami sheets in different handedness. [91] e. Twenty four AuNPs are assembled in a helical fashion along an origami ring to form a LH or RH plasmonic toroidal structure. [90] f. Schematic of the bifacial DNA origami-directed assembly of 3D AuNR dimers. [92] g. Schematic of the self-assembly of RH-AuNR helices. This design enables one-pot assembly of AuNR helical superstructures. [113]

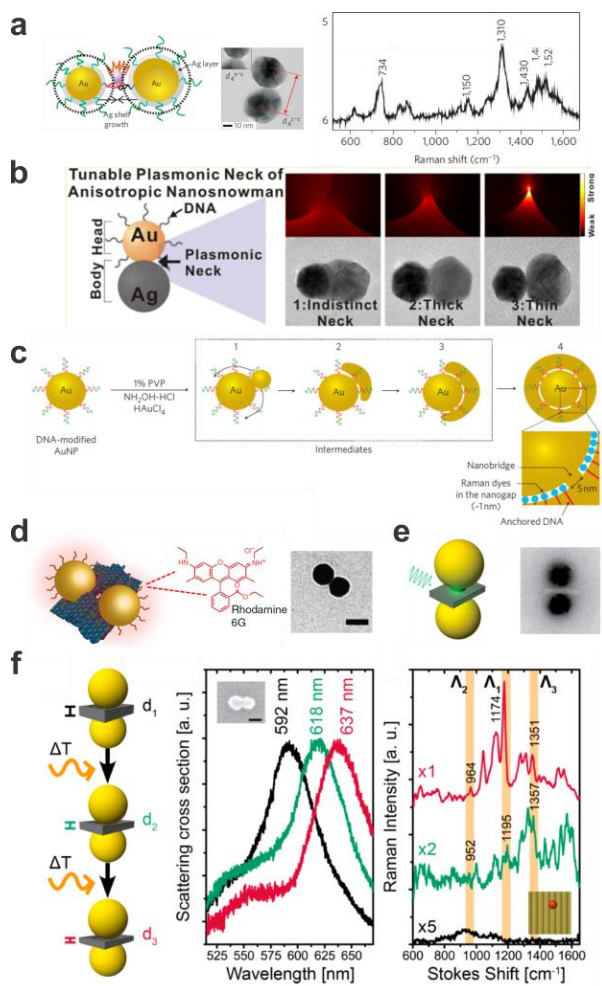

**Figure 6.** Surface-enhanced Raman scattering from DNA-assembled plasmonic nanostructures. a. Ag-shell growth-based AuNP heterodimers using DNA modification. A single Raman-active Cy3 dye molecule is located between DNA-tethered AuNPs. TEM image of a representative structure and SERS spectrum taken from the sample showing characteristic Raman peaks for the Cy3 dye. [27] b. Schematic, calculated electromagnetic field distributions, and TEM images of the Au-Ag nanosnowman structures with various neck junctions. [125] c. Synthetic scheme for the Au nanobridged particles using DNA-modified AuNPs as templates. Raman dyes are enclosed in the nanogaps. [27] d. Schematic of the AuNP dimer assembled on a DNA origami platform with a thin layer of Rhodamine 6G adsorbed onto the dimer structure. TEM image of a representative structure. [30] e. Schematic illustration and TEM image of a DNA origami assembled AuNP hybrid structure used for SERS measurements. A plasmonic hot spot is formed in the gap between the AuNPs for enhancing Raman signals. [32] f. Rayleigh and Raman scattering spectra of an individual AuNP dimer modified with a single Cy3.5 molecule placed at the hot spot before and after a first and a second round of laser excitation. [93]

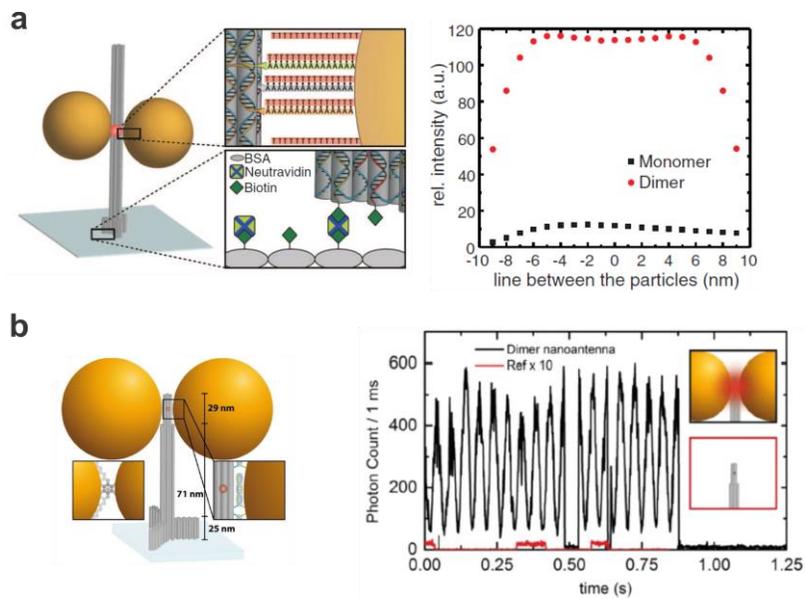

**Figure 7.** Surface-enhanced fluorescence from DNA-assembled plasmonic nanostructures. a. Left: schematic of a DNA origami pillar with two AuNPs forming a dimer. The dye is located between the NPs within the central bundle of the pillar. The DNA origami pillar is bound via biotins to a neutravidin-functionalized cover slip. Right: Numerical simulations of the fluorescence enhancement along the gap for a dye oriented in the radial direction. [36] b. Single-molecule fluorescence transients for a AuNP dimer (black line) and for a DNA origami structure without NPs (red line). [39]

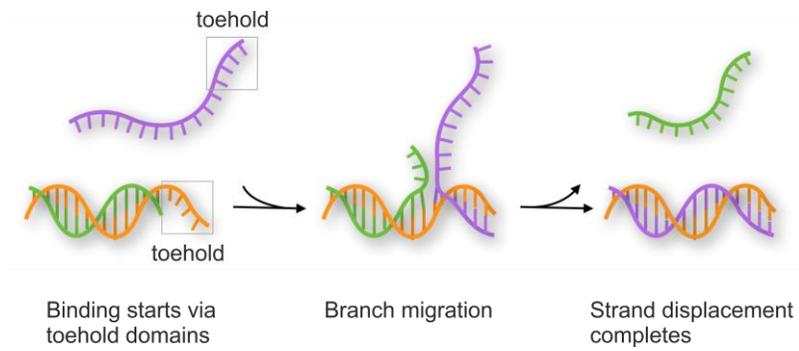

**Figure 8.** Scheme of the toehold-mediated strand displacement process. Note that after binding of the purple strand to the branch can migrate randomly in both directions. This migration starts all over again when only the toehold region is bound but terminates once the branch has reached the left side and the purple strand has fully replaced the green strand.

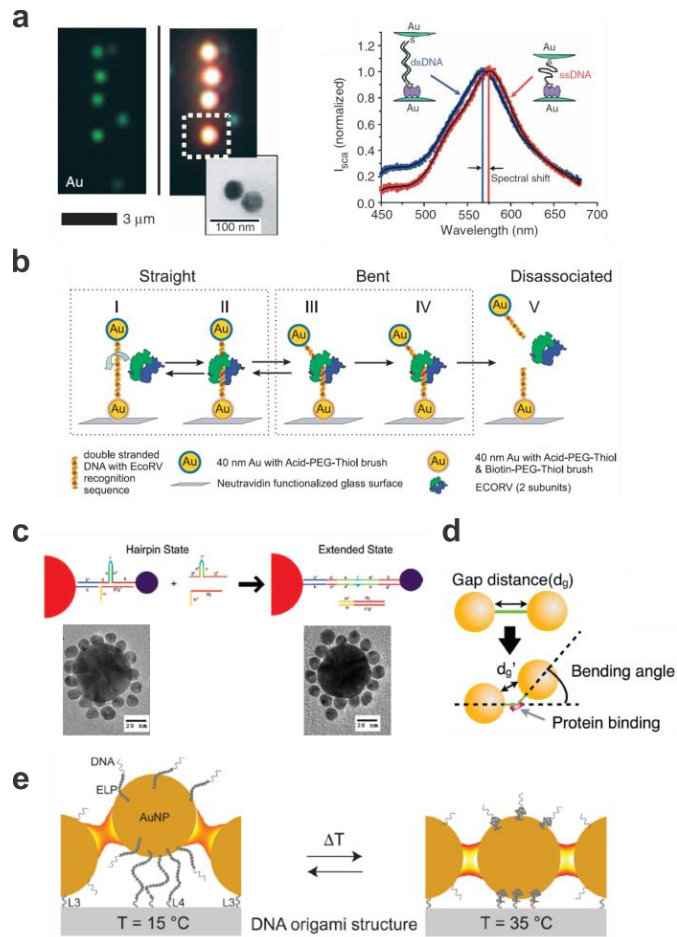

**Figure 9** Plasmon rulers. a. Left: single AuNPs and DNA-assembled AuNP dimers appear blue and blue-green under dark field microscopy, respectively. Right: spectral shift of a AuNP dimer upon DNA hybridization. [22] b. A plasmon ruler is immobilized with one particle to a glass surface through biotin-Neutravidin chemistry. The EcoRV enzyme binds to the DNA strand between the two AuNPs. It bends the DNA strand at the target site and then cuts it creating a blunt ended by phosphoryl transfer. [139] c. Satellite particles are initially linked to a Au core particle with DNA strands incorporated with hairpin bridging strands. Reconfiguration occurs upon addition of strands complementary to the hairpin strands. Corresponding TEM images are shown for the two configurations. [140] d. AuNP dimer before and after DNA bending induced by SOX2. [141] e. Thermoresponsive switching of a plasmonic waveguide. With a temperature increase above the transition temperature, a thermo-sensitive polymer (ELP) can be transformed into the collapsed state leading to shortening of the particle spacing. [142]

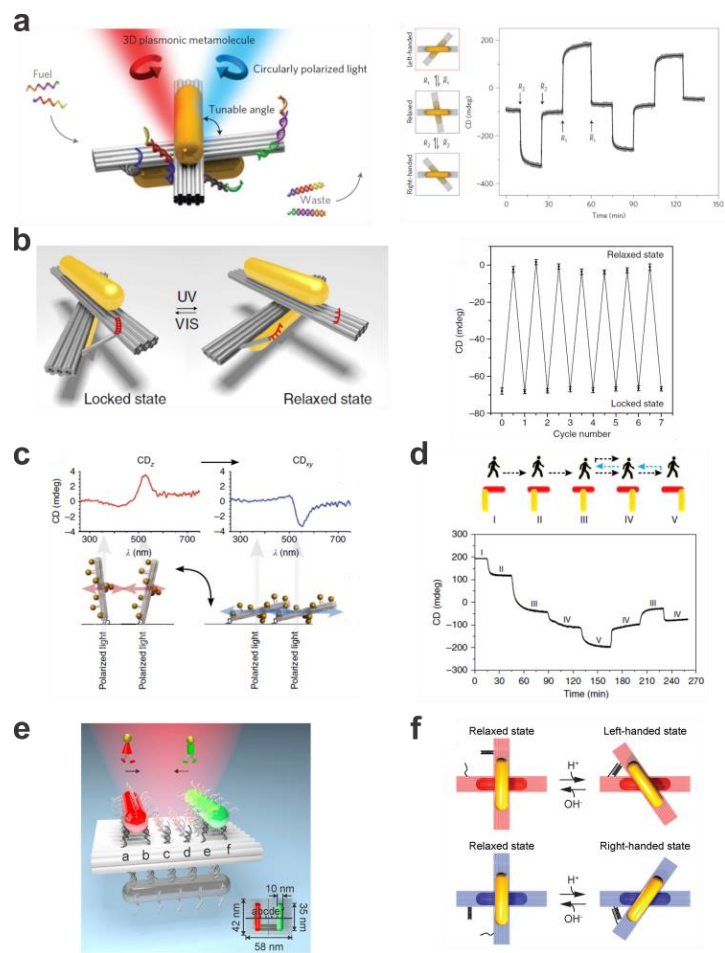

**Figure 10.** DNA-assembled plasmonic nanostructures for dynamic manipulation of strong chirality. a. Reconfigurable 3D plasmonic nanostructures consist of AuNRs hosted on switchable DNA origami templates. The relative angle between the AuNRs within the structure can be controlled using DNA locks, therefore giving rise to dynamic plasmonic chiral responses. [89] b. Light-driven 3D plasmonic nanosystem reversibly regulated by UV and visible light for switching between a right-handed and a relaxed state. The dynamic function of the origami structure is enabled by introducing the azobenzene-modified DNA segment on the origami template. [143] c. Chiral response switched by changing the helix orientation with respect to the light beam. [88] d. A plasmonic walker that can perform stepwise walking on origami. The walking of the AuNR is enabled through toehold-mediated strand displacement. [144] e. A plasmonic walker couple system, in which the two AuNR walkers can independently or simultaneously perform stepwise walking along the same DNA origami track. [145] f. pH regulation of DNA origami-based plasmonic chiral nanostructures. The plasmonic system can be switched between the relaxed and LH/RH state by opening/closing the pH-triggered DNA locks. [146]

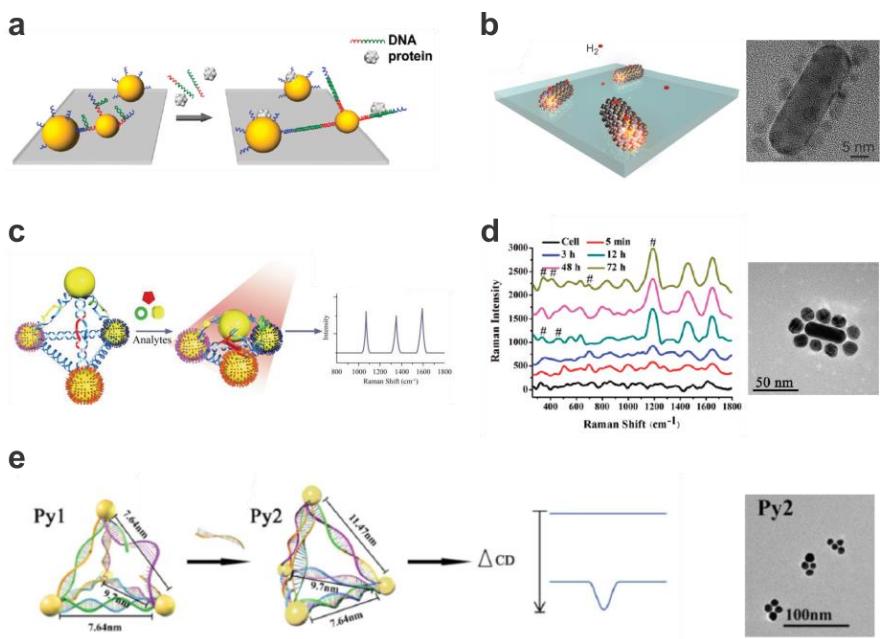

**Figure 11.** Dynamic plasmonic nanostructures for sensing applications. a. DNA-assembled AuNP dimer structures for identifying targets from nonspecific binding and for detecting targets in complex media. [147] b. DNA-assembled bimetallic hydrogen sensors, which consist of palladium nanoparticles linked to AuNRs using DNA strands. TEM image of a representative structure. [148] c. SERS-encoded Ag-pyramids for detection of multiple biomarkers (PSA. thrombin, and mucin-1). [149] d. DNA-assembled plasmonic superstructures composed of AuNRs and spherical AuNPs for SERS-based *in situ* monitoring of intracellular metabolism. TEM image of a representative structure. [150] e. Chiral plasmonic sensors for DNA detection through CD response changes. TEM image of the plasmonic structures. [151]